\documentclass[aps,prl,reprint,superscriptaddress,nofootinbib,preprintnumbers]{revtex4-2}

\usepackage{comment}
\usepackage[utf8]{inputenc} 
\usepackage{graphicx,overpic,mathtools}
\usepackage{amsthm,amsmath,amssymb,hyperref,mathrsfs}
\usepackage{braket,bm,bbm,setspace}
\usepackage[normalem]{ulem} 
\usepackage{physics}
\usepackage[makeroom]{cancel}
\hypersetup{
	colorlinks=true,
	linkcolor=blue,
	filecolor=magenta,      
	urlcolor=cyan,
}
\usepackage[usenames,dvipsnames]{xcolor}

\newcommand{\bfx}{\textbf{x}}

\newcommand{\prop}{\textcolor{blue}}
\newcommand{\luis}{\color{purple}}

\begin{document}

\title{Reply to Comment on “No Black Holes from Light” [\href{https://arxiv.org/abs/2408.06714}{arXiv:2408.06714}]}

\author{\'Alvaro \'Alvarez-Dom\'inguez}
\email{alvalv04@ucm.es}
\affiliation{Departamento de F\'{\i}sica Te\'orica and IPARCOS, Universidad Complutense de Madrid, Plaza de las Ciencias 1, 28040 Madrid, Spain}   

\author{Luis J. Garay} 
\email{luisj.garay@ucm.es}
\affiliation{Departamento de F\'{\i}sica Te\'orica and IPARCOS, Universidad Complutense de Madrid, Plaza de las Ciencias 1, 28040 Madrid, Spain}

\author{Eduardo Mart\'in-Mart\'inez}
\email{emartinmartinez@uwaterloo.ca}
\affiliation{Department of Applied Mathematics, University of Waterloo, Waterloo, Ontario, N2L 3G1, Canada}
\affiliation{Institute for Quantum Computing, University of Waterloo, Waterloo, Ontario, N2L 3G1, Canada}
\affiliation{Perimeter Institute for Theoretical Physics, Waterloo, Ontario, N2L 2Y5, Canada}

\author{Jos\'e Polo-G\'omez}
\email{jpologomez@uwaterloo.ca}
\affiliation{Department of Applied Mathematics, University of Waterloo, Waterloo, Ontario, N2L 3G1, Canada}
\affiliation{Institute for Quantum Computing, University of Waterloo, Waterloo, Ontario, N2L 3G1, Canada}
\affiliation{Perimeter Institute for Theoretical Physics, Waterloo, Ontario, N2L 2Y5, Canada}

\begin{abstract}
We discuss how the comment by A. Loeb [\href{https://arxiv.org/abs/2408.06714}{arXiv:2408.06714}] has no bearing on the results of \href{https://link.aps.org/doi/10.1103/PhysRevLett.133.041401}{Phys. Rev. Lett. \textbf{133}, 041401 (2024)} [\href{https://arxiv.org/abs/2405.02389}{arXiv:2405.02389}].
\end{abstract}

\maketitle



\noindent\textit{\textbf{Introduction and context.---}} In the original letter~\cite{kugelblitz}, we showed that the quantum effects that result from the self-interaction of the electromagnetic field prevent the formation of black holes from the collapse of pure electromagnetic radiation---also known as \textit{kugelblitze}---within the conditions that can be achieved both in a laboratory or in present-day astrophysical scenarios. More concretely, the primary goal of~\cite{kugelblitz} was to demonstrate that quantum effects, particularly the Schwinger effect, act as dissipation sources in the process of concentrating enough electromagnetic energy to create an event horizon. The conclusion was that kugelblitze are impossible in the conditions of our current universe, for radii ranging from $10^{-29}$ to $10^8$ m.  

A recent comment by A. Loeb~\cite{CommentLoeb} incorrectly argues that our study~\cite{kugelblitz} ignores gravitational effects, claiming that this led to a wrong conclusion. 
The comment concludes that black holes can be formed from light by considering gravitational effects 1) in the context of the early universe, where radiation and relativistic particles could lead to black hole formation due to radiation overdensities, and 2) in the case of very small energy densities distributed in a large enough volume.

In this reply, we clarify that gravitational effects were indeed considered in~\cite{kugelblitz}, and that the counterarguments in~\cite{CommentLoeb} do not apply to the original results and therefore do not undermine our conclusions in any way.

\noindent\textit{\textbf{The role of gravitational effects in ~\cite{kugelblitz}.---}} We would like to draw attention to the fact that, despite the claims made by A. Loeb in~\cite{CommentLoeb}, gravitational effects were not ignored in our original work~\cite{kugelblitz}. Rather, their role was explicitly addressed in the point ``Assumption of a Minkowski background'' of the section ``Validity of the results'' of the letter~\cite{kugelblitz}. There, we showed that, in the wide regime of validity of our results, gravitational effects were subleading to dissipative effects, hence justifying the flat spacetime approximation. In these scenarios, the dissipation of energy through pair production ensures that the electromagnetic field never reaches any energy concentration that could lead to spacetime curvatures significant enough to require considering a departure from flat spacetime.




\noindent\textit{\textbf{Early universe criticism.---}} One of the scenarios considered in~\cite{CommentLoeb} is the potential for black hole formation in the early universe. As mentioned explicitly in the section ``Validity of the results'' of the original letter~\cite{kugelblitz}, our study does not apply to the extreme conditions of the early universe. Instead, \cite{kugelblitz} analyzes the possibility of forming kugelblitze in the conditions of the present-day universe, where the Schwinger effect dominates before gravitational effects can play a significant role. The early universe conditions referenced in~\cite{CommentLoeb} involve different dynamics, and our conclusions about the impossibility of forming black holes from light are not affected. 

\noindent\textit{\textbf{Large black hole criticism.---}} The comment~\cite{CommentLoeb} also mentions the possibility of forming black holes that are large enough that they can be formed with radiation below the Schwinger limit, i.e., where particle-antiparticle creation is exponentially suppressed. 
However, the comment fails to acknowledge that the conclusions of~\cite{kugelblitz} specifically pertain to black holes within the already mentioned radius range of $10^{-29}$ to $10^8$ m, where the Schwinger effect is significant. The upper limit of $10^8$~m was explicitly calculated in the section ``Validity of results'' of ~\cite{kugelblitz}, it being precisely a consequence of the fact that above this size, the first inequality of Eq.~(3) of~\cite{kugelblitz} is not guaranteed to be fulfilled, i.e., the electric fields involved are not necessarily above the Schwinger limit. The example provided by A. Loeb in~\cite{CommentLoeb}, where the present-day universe is filled with thermal photons, clearly falls almost twenty orders of magnitude outside of the range of sizes that our study precludes. 


\noindent\textit{\textbf{A back-of-the-envelope calculation.---}} To give some additional intuition on the competition between the attractive effect of gravity and the dissipation via Schwinger effect that prevents the formation of kugelblitze, we offer the following back-of-the-envelope estimation that illustrates how the effects scale with the size of the (potential) black hole. This is not intended to replace the rigorous analysis of the original work, but rather to provide an intuitive understanding of why the Schwinger effect dominates over gravitational effects for the aforementioned window of black hole sizes, thereby preventing the formation of kugelblitze. This simplified approach can help grasp the basic core of the phenomenon without requiring a deep dive into the detailed calculations.

The energy density of a pair produced by Schwinger effect in a strong homogeneous electric field is given by Eq.~(4) of~\cite{kugelblitz}, with the energy density being proportional to the third power of the field strength~$E$. The Supplemental Material of~\cite{kugelblitz} contains a rigorous and detailed calculation of the renormalized stress-energy tensor, which gives the precise result of Eq.~(4) of~\cite{kugelblitz}, but seeing that this is the correct scaling is easy: The standard calculation for the particle density creation rate via Schwinger effect above the Schwinger limit~\cite{Schwinger1951} gives $\text{d}n/\text{d}t\propto E^2$. On the other hand, the energy of each particle in the strong field limit behaves as $\omega\sim \tau_\text{x}E$, where $\tau_\text{x}$ represents the effective time each particle is subjected to the electric field. Therefore, the power is proportional to~$\tau_\text{x}E^3$, and the energy is proportional to $\tau_\text{x}^2E^3$, which is the scaling that the energy density dissipated via the Schwinger effect in Eq.~(4) of~\cite{kugelblitz} shows.  


Now, we can rewrite this scaling in terms of the radius $R$ and the total electromagnetic energy \mbox{$\epsilon=(4\pi R^3/3) \varepsilon_0 E^2/2$}. Since the time that it takes for the (ultrarelativistic)  particles to leave the region is \mbox{$\tau_\text{x}\approx R/c$}, the energy density dissipated scales as
\begin{equation}\label{Eq: 2}
    \lim_{\tau\rightarrow \infty} \langle\hat{T}^{00}(\tau,\bfx)\rangle_{\text{ren}} \sim \tau_\text{x}^2E^3 \sim \epsilon^{\frac32}R^{-\frac{5}{2}}.
\end{equation}
This provides intuition as to why in our calculations the energy density lost via Schwinger effect scales with a negative power of the radius. Now notice that in this estimation there is no a-priori relationship between the total energy $\epsilon$ and the radius, as there would be in a black hole, for which $R=2GM/c^2$. If we want to compare with the mass scaling of the effective energy density of a black hole, we could look at a situation where we have electromagnetic radiation at a point just about enough to produce a black hole of mass~$M$ and just naively take $\epsilon\sim Mc^2$, $R\sim 2G M/c^2$, and then one sees from Eq.~\eqref{Eq: 2} that the Schwinger effect decays slower ($M^{-1}$) than the effective energy density for a black hole ($M^{-2}$) as the mass increases.  This shows that as long as one is past the Schwinger limit there are regimes (precisely the aforementioned range between $10^{-29}$ and $10^8$ m) where the energy density dissipated by the Schwinger effect scales favourably as compared to the energy density necessary to form a black hole, and can therefore overpower gravitational confinement.

\noindent\textit{\textbf{Conclusion.---}}  The arguments presented in the comment~\cite{CommentLoeb} do not apply to the regimes analyzed in~\cite{kugelblitz}. Our study showed that in present-day universe conditions, within the size ranges considered, black holes cannot be formed from light due to quantum effects that prevent energy from concentrating sufficiently to form an event horizon. The flat spacetime approximation used in our calculations is validated within this regime. The criticisms raised in~\cite{CommentLoeb} are not relevant to the claims made in~\cite{kugelblitz}, and therefore do not affect the validity of our original conclusions.


\acknowledgments

AAD and LJG acknowledge  support through the Grant PID2020-118159GB-C44 (funded by MCIN/AEI/10.13039/501100011033). EMM acknowledges support through the Discovery Grant Program of the Natural Sciences and Engineering Research Council of Canada (NSERC). EMM also acknowledges support of his Ontario Early Researcher award. JPG acknowledges the support of a Mike and Ophelia Lazaridis Fellowship, as well as the support of a fellowship from ``La Caixa'' Foundation (ID 100010434, code LCF/BQ/AA20/11820043). Research at Perimeter Institute is supported in part by the Government of Canada through the Department of Innovation, Science and Industry Canada and by the Province of Ontario through the Ministry of Colleges and Universities.


\twocolumngrid

\bibliography{BibKugelblitz}

\end{document}